\documentclass[twocolumn,aps,showpacs,superscriptaddress]{revtex4}
\usepackage{graphicx}
\usepackage{dcolumn}
\usepackage{bm}
\begin{document}
\def\Journal#1#2#3#4{{#1} {\bf #2}, #3 (#4)}

\def\NCA{Nuovo Cimento}
\def\NIM{Nucl. Instr. Meth.}
\def\NIMA{{Nucl. Instr. Meth.} A}
\def\NPB{{Nucl. Phys.} B}
\def\NPA{{Nucl. Phys.} A}
\def\PLB{{Phys. Lett.}  B}
\def\PRL{Phys. Rev. Lett.}
\def\PRC{{Phys. Rev.} C}
\def\PRD{{Phys. Rev.} D}
\def\ZPC{{Z. Phys.} C}
\def\JPG{{J. Phys.} G}
\def\CPC{Comput. Phys. Commun.}
\def\EPJ{{Eur. Phys. J.} C}
\def\PR{Phys. Rept.}

\preprint{}
\title{RHIC physics overview}

\author{Lijuan Ruan}
\affiliation{Physics Department, Brookhaven National Laboratory,
Upton, NY 11973, USA \\
E-mail: ruanlj@rcf.rhic.bnl.gov, ruan@bnl.gov }

\date{\today}
\begin{abstract}
The results from data taken during the last several years at the
Relativistic Heavy-Ion Collider (RHIC) will be reviewed in the
paper. Several selected topics that further our understanding of
constituent quark scaling, jet quenching and color screening
effect of heavy quarkonia in the hot dense medium will be
presented. Detector upgrades will further probe the properties of
Quark Gluon Plasma. Future measurements with upgraded detectors
will be presented. The discovery perspectives from future
measurements will also be discussed.

\end{abstract}
 \maketitle

\section{Introduction}

Data taken in the last few years have demonstrated that the
Relativistic Heavy Ion Collider (RHIC) has created a strongly
interacting hot, dense medium with partonic degrees of freedom,
the Quark Gluon Plasma (QGP) in central Au+Au collisions at
$\sqrt{s_{NN}}$ = 200
GeV~\cite{starwhitepaper,brahmswhitepaper,phenixwhitepaper,phoboswhitepaper}.
Such matter is believed to have existed a few microseconds after
the big bang. Understanding the properties of this matter, such as
the colored degrees of freedom and the equation of state is the
physics goal of RHIC and of broad interest. I will review the
experimental results which were used to identify the existence of
the hot, dense medium, followed by the measurements of its
properties. Future upgrades that are essential to understand the
fundamental properties of the medium will be discussed as well.

RHIC at Brookhaven National Laboratory is the first hadron
accelerator and collider consisting of two independent rings. It
is designed to operate at high collision luminosity over a wide
range of beam energies and particle species ranging from polarized
proton to heavy ion~\cite{rhic:01,rhic:02}, where the top energy
of the colliding center-of-mass energy per nucleon-nucleon pair is
$\sqrt{s_{NN}}$ = 200 GeV. The RHIC facility consists of two rings
with super-conducting magnets, each with a circumference of 3.8
km, which focus and guide the beams. There are four experiments at
RHIC, BRAHMS~\cite{brahmsNIMA}, PHENIX~\cite{phenixEXP},
PHOBOS~\cite{phobosNIMA}, and STAR~\cite{starNIMA}. BRAHMS and
PHOBOS are relatively small experiments. They have finished their
experimental program and were decommissioned after year 2008. STAR
and PHENIX, the two large detectors, are still in operation.

\section{The hot and dense medium created at RHIC}

In 2000, the experiments started to take data. In 2005, the four
experiments published white papers summarizing what has been
discovered. The details can be found in all four white
papers~\cite{starwhitepaper,brahmswhitepaper,phenixwhitepaper,phoboswhitepaper}.
The focus of RHIC was to identify the existence of the QGP. Below
I will review two probes used to identify QGP, bulk probes and
penetrating probes. Bulk probes include measurements of the
majority of produced particles at low $p_T$ ($p_T\!<\!2$ GeV/$c$)
to address the energy density, collectivity and freeze out
properties of hot, dense medium. Penetrating probes are the
measurements of the rarely produced particles such as heavy
flavor, jets and identified particles at high $p_T$ ($p_T\!>\!6$
GeV/$c$) to see the medium effect on their productions and thereby
are used to deduce medium properties. The measurements at
intermediate $p_T$ ($2\!<\!p_T\!<\!6$ GeV/$c$) probe the interplay
between bulk and hard components and reveal some unique
interesting features of the collisions at RHIC.

\subsection{Bulk properties}
Many measurements at RHIC have studied the bulk properties of the
collisions, including low $p_T$ identified particle transverse
momentum and rapidity distributions in different collision
centralities, system sizes and collision energies. The
measurements indicate that RHIC has created a hot and dense
partonic medium which expands and cools down hydrodynamically.
Hadrons freeze out chemically at close to critical temperature
($T_c$) and then freeze out kinetically at lower temperature.
Below are several important measurements that point to the
existence of a hot, dense medium and its freeze out features.

\begin{itemize}
\item The rapidity dependence of particle multiplicity
demonstrates that the 26 TeV energy has been dumped in the system
to produce particles in 200 GeV central Au+Au
collisions~\cite{brahms04energydensity}. The energy density is
much higher than normal nuclear matter density thus it is believed
that partonic matter is formed in such
collisions~\cite{phenixenergydensity}.

\item The measurements of elliptic flow $v_2$, the second harmonic
coefficient of a Fourier expansion of the final momentum-space
anisotropic azimuthal distribution, show mass dependence at low
$p_T$ consistent with hydrodynamic behaviors with quark gluon
plasma equation of state~\cite{starv2,hydro}. The quantitative
consistency between data and hydrodynamic model calculations is
lacked and complicated by initial conditions, hadronization and
freeze out processes. However, the magnitude of the elliptic low
can be reproduced only when partonic interactions are included in
the calculations.

\item The identified particle $p_T$ distributions were measured
and fit with the blast-wave model using thermal-like distribution
or Tsallis-like distributions~\cite{starPID:04,Blastwave,Tsallis}.
The latter include the fluctuation and non-equilibrium effects in
the system and can be applied in p+p collisions also. The
thermal-like fit indicates that the kinetic freeze out temperature
$T_{kin}$ decreases from p+p to peripheral Au+Au to central Au+Au
collisions while the velocity profile increases. This indicates
that the system is cooling with expansion.

\item The identified particle ratios measured in Au+Au collisions
at different centralities were fit with thermal model
distributions~\cite{starPID:04,thermalmodel}. The fit indicates
that the chemical freeze out temperature ($T_{chemical}$) is about
160 MeV and that there is no significant centrality or system size
dependence. This value of $T_{chemical}$ is very close to the
critical temperature $T_c$ calculated by Lattice
QCD~\cite{latticeTc}.

\item The measurements of resonance to stable particle ratios and
Hanbury Brown-Twiss (HBT) interferometry of two particle
Bose-Einstein correlations indicate that the time interval between
chemical and kinetic freeze out is about 3-10
fm/c~\cite{K*04,hbt}.
\end{itemize}

The above measurements are consistent with the physics picture in
which partonic matter is created at RHIC, expands and cools down,
and hadronizes into a system of hadrons that freeze out chemically
shortly after at a fixed temperature, followed by further
expansion and cooling down until kinetic freeze-out.

\subsection {Hard probes and medium properties}

Beside bulk probes, hard probes such as identified particles at
high $p_T$, jets and and heavy flavor are thought to be ideal
probes for quark gluon plasma. They are thought to be well
calibrated since they are believed to be produced from hard
processes with high $Q^2$ transfer and thus can be calculated in
perturbative Quantum Chromodynamic (pQCD) framework~\cite{pQCD}.
In pQCD calculations, identified particle production can be
described as a convolution of parton distribution functions,
parton parton interaction cross sections and parton fragmentation
functions. When hard partons traverse the hot and dense medium
created in the collision, they lose energy by gluon radiation
and/or colliding elastically with surrounding
partons~\cite{jetquench,starhighpt,rhicotherhighpt}. This
phenomenon is also called jet quenching. Jet quenching leads to a
softening of the final measured hadron spectra at high $p_T$. The
amount of energy loss can be calculated in QCD and is expected to
be different for energetic gluons, light quarks and heavy
quarks~\cite{xinnian:98,deadcone}. In experiments, in order to
quantify the effect of the medium, the nuclear modification
factors ($R_{AB}$ or $R_{CP}$) are measured, where the invariant
yield in A+B collisions is divided by that in p+p or peripheral
A+B collisions, scaled by their respective numbers of binary
nucleon-nucleon collisions. If there was no nuclear medium effect,
the ratio would be 1 at high $p_T$. Any deviation from unity
therefore indicates nuclear medium effects. This is similar to
medical imaging techniques, where the picture of a human body can
be obtained through the calibrated gamma ray interaction. Below
several important results are listed to illustrate the properties
of the medium through the change in hard probes in central
heavy-ion collisions compared to peripheral and to p+p collisions.

\begin{itemize}
\item The $R_{AA}$ of inclusive hadrons in central Au+Au
collisions at 200 GeV at mid-rapidity shows a factor of 5
suppression with respective to unity at $p_{T}\!>\!6$
GeV/c~\cite{starhighpt,rhicotherhighpt}. The pQCD calculation with
gluon density $dN_{g}/dy=1000$ and with radiative energy loss can
describe the suppression~\cite{jetquench}. The $R_{dAu}$ of
inclusive charged hadrons in d+Au collisions shows enhancement at
intermediate $p_T$ and equals to unity at high
$p_T$~\cite{rhicdau}. This indicates that the strong suppression
observed in $R_{AA}$ in central Au+Au collisions is due to final
state effects and not due to an initial wave function difference
such as a possible color glass condensate (CGC)~\cite{CGC} at
mid-rapidity.

\item Two particle azimuthal angle correlations show that in
central Au+Au collisions, away side particle production at
$p_{T}\!>\!2$ GeV/c disappears or is suppressed significantly
compared to in p+p and d+Au collisions with respect to a high
$p_T$ trigger ($p_T\!>\!6$
GeV/c)~\cite{starwhitepaper,rhiccorrelation}. When the $p_T$ was
lowered for associate particles, enhancement of particle
production on the away side was observed compared to p+p and d+Au
collisions, and the transverse momentum distribution on the away
side is softened and approaches the inclusive particle
distribution~\cite{rhiccorrelationII}. This indicates that the
energy loss by the jet on the away side might be thermalized by
the system.
\end{itemize}

The above measurements indicate that the suppression on $R_{AA}$
of high $p_T$ particles in central Au+Au collisions are consistent
with partonic energy loss picture.

\subsection {Intermediate $p_T$ physics: number of constituent quark
scaling (NCQ) and baryon enhancement}

Between low $p_T$ where the physics is dominated by bulk
properties and high $p_T$ where the particle production is by jet
fragmentation, there is also rich physics which can be used to
explore the properties of the medium created in heavy ion
collisions. Below several interesting results are presented.

\begin{itemize}
\item The identified particle elliptic flow measurements for
$\pi$, K, $p$, $\Xi$, $\Omega$ and $\phi$ indicate that the flow
pattern at intermediate $p_T$ seems to follow a simple scaling
governed by the fact that mesons (baryons) has two (three)
constituent quarks~\cite{NCQrhicv2}. Even though multi-strange
hadrons or $\phi$ have smaller interaction cross sections at
hadronic stage, they have a similar flow pattern as non-strange
hadrons. This indicates that the elliptic flow is mainly developed
at the partonic stage where the light-strange quark difference is
insignificant. Coalescence or recombination
models~\cite{recombinemodels}, in which two or three constituent
quarks are combined into mesons or baryons, were proposed to
explain the data.

\item At intermediate $p_T$, $R_{CP}$ ($R_{AA}$) for baryons is
larger than that for mesons, indicating strong baryon enhancement
in Au+Au collisions~\cite{baryonenhancement}. In central Au+Au
collisions, the $p/\pi$ ratio reaches unity, which is much larger
than that from elementary p+p collisions. The coalescence or
recombination model can qualitatively reproduce the feature. The
parton density at RHIC is significant so that parton recombination
into hadrons is efficient. In the same $p_T$ region, the parton
$p_T$ for baryons is effectively lower than that for mesons, thus
the baryon over meson ratio can be significantly enhanced in the
intermediate $p_T$ region in central Au+Au collisions.
\end{itemize}

At intermediate $p_T$, elliptic flow and baryon over meson ratio
measurements are consistent with the recombination or coalescence
picture in which partons recombine into hadrons at hadronization.

To summarize this section, the measurements on bulk properties,
hard penetrating probes and at intermediate $p_T$ at RHIC indicate
that RHIC has created a dense and rapidly thermalizing matter
characterized by: 1) initial energy densities far above the
critical values predicted by lattice QCD for formation of a QGP;
2) opacity to jets; and 3) nearly ideal fluid flow, which is
marked by constituent interactions of very short mean free path,
established most probably at a stage preceding hadron
formation~\cite{starwhitepaper}.

The next objective is to study the properties of the created
matter in detail in terms of the equation of state and colored
degrees of freedom. For example, one would like to know the
temperature, the chemical composition and the velocity of sound of
the hot and dense matter. One would also like to further
understand or test jet quenching and NCQ scaling and to study
other signatures of the quark gluon plasma such as color screening
effects.

\section{several selected recent highlights from RHIC}

In this section, several selected recent results will be presented
that further our understanding of partonic flow, NCQ scaling and
jet quenching. New results on possible color screening effects of
QGP are also presented.

\subsection{Further measurements of partonic flow and NCQ scaling}
\begin{figure}
\includegraphics*[keepaspectratio,scale=0.45]{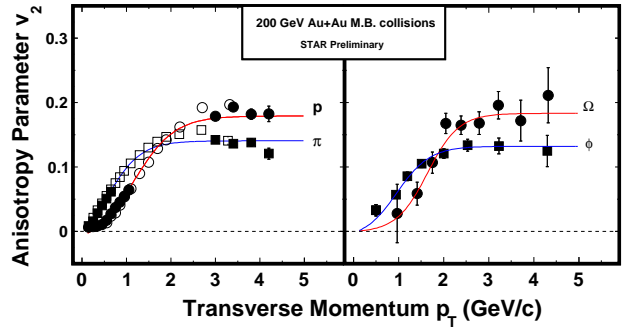}
\caption{$v_{2}$ as a function of $p_T$ for $\pi$, $p$, $\Omega$
and $\phi$ in 200 GeV minimum bias Au+Au collisions. Open symbols
are from PHENIX~\cite{phenixflow} and the solid symbols are from
STAR. Lines represent NCQ-inspired fit~\cite{xdong2004}. The
figure is taken from ref.~\cite{shusu:09}.} \label{Figure1}
\end{figure}

With the high statistics data from Au+Au collisions taken in year
2007, STAR measured $v_2$ of $\phi$ and $\Omega$ with high
precision, shown in Fig.~\ref{Figure1}. The $\Omega$ and $\phi$
$v_2$ at intermediate $p_T$ is close to the proton and pion $v_2$
respectively. This indicates that $v_2$ observed at RHIC is
dominantly due to partonic collectivity~\cite{shusu:09}. With high
precision measurements, STAR and PHENIX both showed that there is
significant deviation from NCQ scaling for $v_2$ measurements at
intermediate $p_T$~\cite{shengli:08,na:09}, which can be
understood as the hard component from jet fragmentation starting
to play a role in the corresponding $p_T$ region.

\subsection{The characteristics of jet quenching}
To further understand energy loss mechanisms and medium
properties, nuclear modification factors for direct photons were
measured. The $R_{AA}$ from direct photons, which are the
inclusive photon yields subtracting hadronic decay contributions,
is consistent with no suppression at high
$p_T$~\cite{directphotonpre}. This confirms that the suppression
observed in the $R_{AA}$ for hadrons is due to jet quenching,
rather than initial wave function change which would affect the
direct photons as well.

In addition, nuclear modification factors for protons, pions,
non-photonic electrons from heavy flavor decay were also measured
to test color charge or flavor dependence of energy loss. For
example, gluons carry different Casmir factor from quarks. The
coupling of gluons to the medium is stronger than the coupling of
quarks to the medium thus gluons are expected to lose more energy
than quarks when traversing the medium. At RHIC energy, the gluon
jet contribution to protons is significantly larger than to pions
at high $p_T$~\cite{xinnian:98,AKK,ppdAuPID}. Therefore, protons
are expected to be more suppressed than pions in $R_{AA}$ or
$R_{CP}$ measurement. Experimentally, protons and pions show
similar magnitudes of suppression in $R_{CP}$~\cite{starAuAuPID}.
One of the proposed mechanisms is the jet conversion
mechanism~\cite{weiliu:07}, in which, a jet can change flavor or
color charge after interaction with the medium. With much larger
jet conversion cross sections compared to that in the Leading
Order (LO) calculation, the proton and pion suppression magnitudes
are similar. Using the same factor scaling the LO QCD
calculations, kaons are predicted to be less suppressed than pions
since the initially produced hard strange quarks are much fewer
than the strange quarks in a hot, dense medium~\cite{Fries:08}.
Alternatively, enhanced parton splitting in the medium will also
lead to a change of the jet hadron chemical composition in Au+Au
collisions compared to that in p+p collisions~\cite{urs:07}.

\begin{figure}
\includegraphics*[keepaspectratio,scale=0.4]{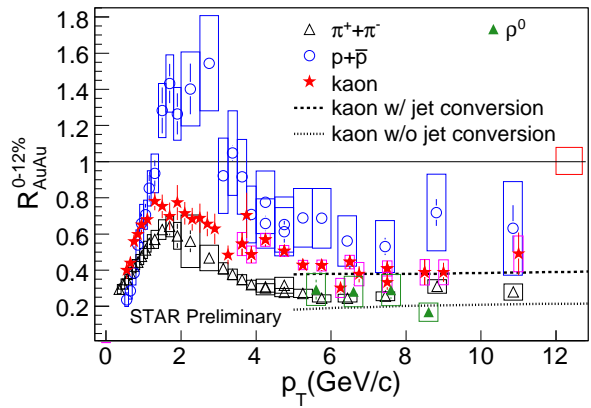}
\caption{Nuclear modification factors of pion, kaon, and proton
and rho in Au+Au collisions. The bars and boxes represent
statistical and systematic uncertainties. The figure is taken from
ref.~\cite{yichun:09}.} \label{Figure2}
\end{figure}

The new measurements of strange hadrons can test further these
mechanisms. Recently, STAR showed the invariant yields of $\pi$, K
and $p$ up to $p_T$ of 15 GeV/c in 200 GeV p+p collisions at
mid-rapidity, which can further constrain light flavor separated
quark and gluon fragmentation functions and serves as a baseline
for the prediction of Au+Au collisions at high $p_T$. The
preliminary results of $R_{AA}$ in central Au+Au collisions
indicated $R_{AA}(K_{S}^{0}, K^{\pm},
p+\bar{p})\!>R_{AA}(\pi^{+}+\pi^{-})\!\sim\! R_{AA}(\rho^{0})$, as
shown in Fig.~\ref{Figure2}~\cite{yichun:09}. This provides
additional constraints on energy loss calculations. A full
comparison between data and calculations requires consideration of
quantitative modelling and calculations incorporating 3D hydro in
an expanding medium~\cite{renk:07} and proper light
flavor-separated quark and gluon fragmentation functions.
Experimentally, high $p_T$ strange hadron measurements in d+Au,
its centrality dependence of $R_{AA}$ in Au+Au and elliptic flow
$v_{2}$ measurements will shed more light on our understanding of
energy loss mechanisms.

On the other hand, non-photonic electrons, which come from heavy
flavor charm and bottom decay, show a similar magnitude of
suppression as light hadrons~\cite{starelectron,phenixelectron}.
The pQCD calculations including collisional and radiative energy
loss show a systematically higher $R_{AA}$ value than experimental
data~\cite{wicks,bdmps}. Further calculations indicate that with
the charm contribution only, non-photonic electrons are expected
to reproduce the data~\cite{bdmps}. Using the azimuthal angle
correlations between non-photonic electrons and charged hadrons
(e-h) and between non-photonic electrons and $D^{0}$ ($e-D^{0}$),
the bottom contribution factor to non-photonic electrons were
measured~\cite{ehcorrelation}. It was found that at $p_T\!>5$
GeV/$c$, the bottom contribution is very significant. This
together with non-photonic electron $R_{AA}$ measurements
challenge the pQCD energy loss model calculations; they may
indicate collisional dissociation of heavy mesons~\cite{vitev:07},
in-medium heavy resonance diffusion~\cite{rapp:06}, and multi-body
mechanisms~\cite{liuko:06} might play an important role for heavy
quark interactions with the medium.

\subsection{Color screening effect on high $p_T$ $J/\psi$?}
The dissociation of quarkonia due to color screening in a QGP is a
classic signature of de-confinement in relativistic heavy-ion
collisions~\cite{colorscreen}. Results at RHIC show that the
suppression of the $J/\psi$ as a function of centrality (the
number of participants) is similar to that observed at the SPS,
even though the energy density reached in collisions at RHIC is
significantly higher~\cite{Adare:2006ns,Abreu:2000xe}. Possible
production mechanisms such as sequential
suppression~\cite{satz_0512217}, $c\bar{c}$
recombination~\cite{BraunMunzinger:2000px,Grandchamp:2001pf,Gorenstein:2001xp,Thews:2000rj}
were proposed to explain this. Recent Lattice QCD calculations
indicate that direct $J/\psi$ is not dissociated in the medium
created at RHIC while the suppression observed for $J/\psi$ comes
from the dissociation of $\chi_{c}$ and
$\psi^{'}$~\cite{latticeqcd}. However, the direct $J/\psi$ might
be dissociated at RHIC at high $p_{T}$, which was predicted in the
hot wind dissociation picture, in which the AdS/CFT approach was
used and the dissociation temperature for $J/\psi$ was predicted
to decrease as a function of $J/\psi$ $p_{T}$~\cite{adscft}. The
AdS/CFT approach was applied to hydro framework and predicted that
$J/\psi$ $R_{AA}$ decreases versus $p_{T}$~\cite{hydro+hotWind}.

\begin{figure}
\includegraphics*[keepaspectratio,scale=0.4]{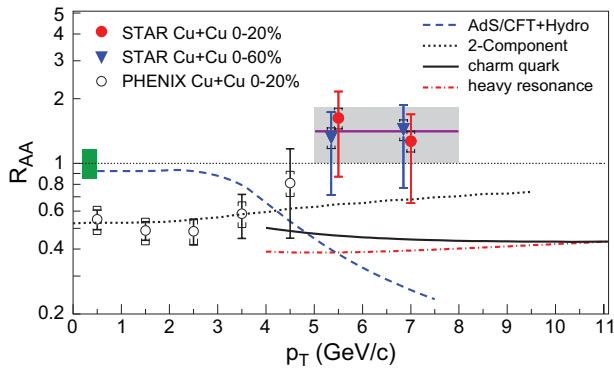}
\caption{$J/\psi$ $R_{AA}$ versus $p_{T}$. The box about unity on
the left shows $R_{AA}$ normalization uncertainty, which is the
sum in quadrature of p+p normalization and binary collision
scaling uncertainties. The solid line and band show the average
and uncertainty of the two 0-20\% data points. The curves are
model calculations described in the text. The uncertainty band of
10\% for the dotted curve is not shown. Figure is taken from
ref.~\cite{starjpsi}.} \label{Figure3}
\end{figure}

Figure~\ref{Figure3} shows $J/\psi$ $R_{AA}$ as a function of
$p_{T}$ in 0-20\% and 0-60\% Cu+Cu collisions from
STAR~\cite{starjpsi} and 0-20\% Cu+Cu collisions from
PHENIX~\cite{phenixjpsi}. The average of two STAR 0-20\% data
points at high $p_T$ is $R_{AA}\!=\!1.4\! \!\pm\! \!0.4
(stat.)\!\pm\! \!0.2 (syst.)$. Compared to low $p_{T}$ PHENIX
measurements, the results indicate that $R_{AA}$ of $J/\psi$
increases from low $p_{T}$ to high $p_T$ at the 97\% confidence
level (C.L.). The $R_{AA}$ of high $p_T$ $J/\psi$ is in contrast
to strong suppression for open
charm~\cite{wicks,vitev:07,charmcucu}, indicating that $J/\psi$
might be dominantly produced through color singlet configuration.
However, even though there is significant improvement from the
next-next-to-leading order (NNLO) pQCD calculations with the color
singlet model, the calculation still fails to reproduce the high
$p_T$ part~\cite{jpsicsNNLO}. The $R_{AA}$ trend of $J/\psi$ is
contradictory to AdS/CFT+hydrodynamic calculations at the 99\%
C.L.. This might indicate two things: 1) Cu+Cu system is not big
enough so that the calculation is not applicable. The larger
system produced in Au+Au collisions may be necessary to observe or
exclude the effect predicted by AdS/CFT; 2) the formation time
effect for high $p_T$ $J/\psi$ is important since the
AdS/CFT+hydrodynamic calculation shown in Fig.~\ref{Figure3}
requires that the $J/\psi$ be produced as an on-shell $J/\psi$
fermion pair, almost instantaneously, at the initial impact with
no formation time. A calculation combining effects of $J/\psi$
formation time, color screening, hadronic phase dissociation,
statistical $c\bar{c}$ coalescence and B meson feed-down
contribution can describe the data~\cite{two_component_approach}.
The calculation suggests a slight increase in the $R_{AA}$ at
higher $p_T$.

In summary, the recent measurements further confirm that a
partonic, hot and dense medium is created in central Au+Au
collisions at RHIC. The strong suppression in $R_{AA}$ for hadrons
is due to jet quenching. However, even though the framework of jet
quenching might be valid, the details of how jets interact with
the medium and lose energy need more detailed theoretical
assessments and coherent modelling is required. To understand
possible color screening effects of quarkonia, it is necessary to
understand their production mechanisms in elementary p+p
collisions. Nuclear modification factors for heavy quarkonia can
help further constrain their production mechanisms as well.

\section{Future upgrades and the related key measurements}

STAR and PHENIX recently updated their data acquisition and
trigger systems, which will help sample RHIC II luminosity. With
several detector upgrades, heavy flavor collectivity and energy
loss, color screening effects, QGP thermal
radiation~\cite{dilepton,dileptonII,rhicIIQuarkonia} and jet
quenching will be studied with better precision at RHIC.

\subsection{Heavy flavor collectivity and energy loss}
The non-photonic electron analyses suffer from big systematic
uncertainties, which are related to photonic background
reconstruction and/or subtraction from hadronic decays. In the
future, with the Heavy Flavor Tracker upgrade at STAR, the direct
topological reconstruction of heavy favor hadron decays will be
feasible and direct charmed hadron measurements will be obtained
with good precision~\cite{hft}. With the Silicon Vertex Detector
upgrades, PHENIX will be able to measure non-photonic electrons
from charm and bottom decay separately~\cite{svt}. These
measurements are crucial to understand heavy flavor energy loss
thus further constrain the details of jet quenching. The
collectivity measurements from heavy flavor will be important to
understand the thermalization for light flavor.

\subsection{Quarkonia production mechanisms, color screening, collectivity and energy loss}
To further understand the production mechanisms of quarkonia,
color screening effects and medium properties, the precise
measurements of the following are needed: nuclear modification
factors of $J/\psi$ from low to high $p_T$ in Au+Au and d+Au
collisions, $J/\psi$ $v_2$, forward and backward $J/\psi$
production to address intrinsic charm contributions at large
$x_F$~\cite{perkins}, $J/\psi\!-\!h$ correlations to access the
feeddown contribution, the spin alignment of
$J/\psi$~\cite{phenixjpsi:09}, higher charmonia states and
different $\Upsilon$ states~\cite{phenixandstar:qm09}. The
$\Upsilon$ states are also ideal tools to study the effect of
color screening in hot and dense QCD matter since its ground state
and excited states melt at different temperatures and all of them
decay to dileptons~\cite{satz_0512217}. Furthermore, since the
$b\bar{b}$ cross section at RHIC energy is expected to be much
smaller compared to $c\bar{c}$  cross section from FONLL
calculations~\cite{vogt}, the recombination contribution from QGP
phase might be negligible to bottomonia production. This makes the
$\Upsilon$ even a better probe for studying the color screening
effect in QGP if sufficient statistics can be achieved
experimentally. The Time of Flight system~\cite{tof}, fully
installed in the summer of 2009, will enhance the $J/\psi$
capability at low $p_T$ significantly at STAR. RHIC II luminosity
enables the $\Upsilon$ $R_{AA}$ measurements with good precision.
With the possible Muon Telescope Detector upgrade, STAR can
cleanly separate the ground state from the excited states even
with the additional material from the upgraded inner tracker since
the muons in $\Upsilon\rightarrow\mu^{+}\mu^{-}$ do not suffer
from Bremsstrahlung radiation~\cite{mtd}. With the Silicon Vertex
Detector upgrade, PHENIX can measure different upsilon states as
well through $\Upsilon\rightarrow e^{+} e^{-}$ since the silicon
vertex detector brings better mass resolution to quarkonia
measurement. The different state $\Upsilon$ measurements will shed
more light on the study of the temperature of the QGP created at
RHIC.

\subsection{Dilepton measurements in the future: vector meson properties and continuum}
\begin{figure*}[t]
 \includegraphics[width=0.625\linewidth]{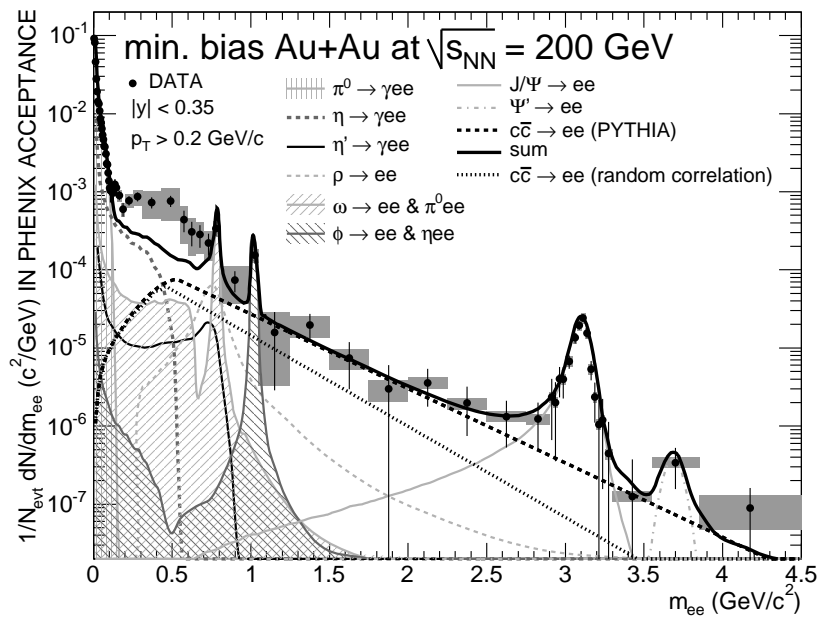}
 \caption{  \label{Figure4}
Invariant e$^+$e$^-$--pair yield compared to the yield from
simulated hadron decays. The charmed meson decay contribution
based on PYTHIA is included in the sum (solid black line). The
charm contribution expected if the dynamic correlation of $c$ and
$\bar{c}$ is removed is shown separately. Statistical (bars) and
systematic (boxes) uncertainties are shown separately; the mass
range covered by each data point is given by horizontal bars. The
systematic uncertainty on the cocktail is not shown. Figure is
taken from ref.~\cite{phenixdilepton}.}
\end{figure*}

The dilepton spectra at intermediate mass range are directly
related to thermal radiation of the
QGP~\cite{dilepton,dileptonII}. At low mass range, we can study
the vector meson in-medium properties through their dilepton
decays, the observable of possible chiral symmetry restoration.
For example, we can measure $\phi\rightarrow e^{+} e^{-} $ and
$\phi\rightarrow K^{+} K^{-} $ in p+p and Au+Au collisions to see
whether the yield ratios from these two decay channels are the
same or not. We can measure $\rho\rightarrow e^{+} e^{-} $ to see
whether there is a mass shift or broadening and also compare the
possible $a_{1}\rightarrow \gamma \pi $ measurements. These
measurements will shed light on the study of chiral symmetry
restoration. At the intermediate mass region, in order to get the
signature of QGP thermal radiation, the $c\bar{c}$ from heavy
flavor decay must be subtracted. Figure~\ref{Figure4} shows the
dilepton invariant mass distribution after background subtraction.
You can see at low mass region, there is significant
enhancement~\cite{phenixdilepton}. Further study indicates that
the enhancement is mainly at low $p_T$. From the low mass and
higher $p_T$ region, the direct photon measurements were obtained
at $1\!<p_T\!<5$ GeV/c. The average temperature of QGP at RHIC was
obtained by PHENIX~\cite{phenixdileptonII}. At the intermediate
mass region, currently, there is no conclusion yet as to whether
there is enhancement from thermal radiation or not. The current
measurement suffered from large systematic uncertainties. In the
future, the precise measurements of D mesons with the Heavy Flavor
Tracker~\cite{hft} and non-photonic electrons from charm decay
with the Silicon Vertex Detector~\cite{svt} will help constrain
the $c\bar{c}$ background contribution. However, the measurement
of $c\bar{c}$ correlation is still challenging. The Muon Telescope
Detector in STAR will provide $\mu\!-\!e$ correlation for the much
needed independent measurements of heavy-flavor contribution to
the dileptons~\cite{mtd}.

\subsection{Direct photons and jet reconstruction}
Direct photons at high $p_T$ are believed to be a golden probe to
study jet quenching since photons do not interact with the
medium~\cite{directphotonI}. For example, if we trigger on a
direct photon and look at away side hadrons or identified
particles, the fragmentation function can be precisely studied and
compared to theoretical calculations~\cite{directphotonII}. Both
STAR and PHENIX measured the fragmentation functions in p+p and
Au+Au collisions with respect to a direct photon trigger and the
results are consistent with most of the theoretical
calculations~\cite{directphoton}. In the future, with RHIC II
luminosity, precise measurements on the away side in both p+p and
Au+Au will shed light on the path length dependence of energy
loss~\cite{directphotonIII}, color charge dependence of energy
loss~\cite{xinnian:98} and jet conversion etc. Recently, the
results for full jet reconstruction at RHIC were also
shown~\cite{fulljet}. The preliminary results indicate a
significant broadening of energy profile in Au+Au collisions. RHIC
II luminosity will make the measurements more precise thus how the
jet energy loss disperses in the medium will be better understood.

To summarize the section, detector upgrades together with RHIC II
luminosity will enable the RHIC experiments to put further
constraints on the characteristics of jet quenching. The study of
color screening through different quarkonia states and precise
dilepton measurements will also be enabled. Possible thermal
radiation signatures may be obtained allowing the temperature of
the QGP created at RHIC to be inferred. While the efforts
mentioned above are mainly to study the properties of QGP in
detail, an effort to understand phase transition and explore the
phase diagram is also on-going at RHIC.

\section{Energy scan: looking for the critical point}
The baryon chemical potential ($\mu_{B}$) at top energy at RHIC is
very close to zero~\cite{starPID:04, starwhitepaper}. At top
energy, the transition from hadronic to partonic matter was
thought to be a cross-over transition~\cite{rhicmuonbI}. When we
go to larger $\mu_{B}$, several calculations indicate that the
transition might be first order~\cite{rhicmuonbII}. To know where
the first order phase transition ends is of great interest. The
signature of a first order phase transition such as long range
fluctuations in event-by-event observables will be measured.
Together with the identified particle ratio measurements, the
$\mu_{B}$ and $T_{chemical}$ freeze out temperature can be
obtained.

At RHIC, a lower energy scan has been proposed to study the phase
diagram and also to see at what energy, the signatures of QGP
disappear such as jet quenching, large elliptic flow and NCQ
scaling etc~\cite{rhicenergyscan}. In 2007, RHIC had a test run at
9.2 GeV. STAR has analyzed the data, 3000 good events, for
identified particle spectra, elliptic flow and HBT radii and
submitted a paper with these results for
publication~\cite{starpid:09}. Compared to a previous lower energy
program at the SPS, RHIC experiments have uniform acceptance for
all beam energies thus systematic uncertainties can be reduced.
Also there will be less ambiguity when comparing the results in
different energies. The large acceptance and excellent particle
identification from the detectors at RHIC will enable significant
qualitative and quantitative improvements in the measurements
compared to SPS.

\section{Discovery possibilities}
Beside the major discovery of QGP created at RHIC, there are many
other discovery possibilities. Recently, the anti-hypertriton was
measured at the STAR experiment. This is the first observation of
an antimatter hypernucleus~\cite{hypertriton}. This opens the
window to studying hyperon-baryon interactions. Also at RHIC, STAR
collaborators found that the correlation between same (different)
charged sign particles is positive (negative)~\cite{starPV}. This
is qualitatively consistent with the strong parity violation
picture, which induces charge separation with respect to the
reaction plane. The charge separation can occur if two possible
scenarios exist: chiral symmetry restoration and a strong magnetic
field in QGP~\cite{dimaPV}. Similar measurements at lower beam
energies will help the understanding of this effect and determine
if strong parity violation is the only explanation for the
observation. In addition, the search for the CGC~\cite{CGC} and
QCD critical point~\cite{rhicenergyscan,bedanga:09} are important
and ongoing programs at RHIC too.

\section{Conclusion}
In summary, I have presented measurements that identify the
existence of QGP at RHIC. Several recent new measurements are
presented that further our understanding of partonic flow and NCQ
scaling, jet quenching and the color screening effect in QGP. The
future upgrades at RHIC will significantly enhance the capability
for dilepton and heavy flavor measurements, which will further our
understanding of the properties of QGP.

\section{Acknowledgements}
The author would like to thank  H.Z. Huang, B. Mohanty, Z. Tang,
Y. Xu and Z. Xu for many valuable discussions. Thank G. Eppley for
proof reading. This work was supported in part by the U. S.
Department of Energy under Contract No. DE-AC02-98CH10886. L. Ruan
is supported in part by the Battelle Memorial Institute and Stony
Brook University in the form of the Gertrude and Maurice Goldhaber
Distinguished Fellowship.

\end{document}